\documentclass{aa}
\usepackage{natbib}
\usepackage{graphicx}
\usepackage{txfonts}
\usepackage{epsfig}
\bibpunct{(}{)}{,}{}{}{,}

\begin{document}
\title{Dynamics of isolated magnetic bright points derived from Hinode/SOT G-band observations}
\titlerunning{Dynamics of isolated magnetic bright points}

\author{D. Utz
\inst 1
\and A. Hanslmeier
\inst {1, 2}
\and R. Muller
\inst 2
\and A. Veronig
\inst 1
\and J. Ryb{\'a}k
\inst 3
\and H. Muthsam
\inst 4}

\institute{IGAM/Institute of Physics, University of Graz, Universit{\"a}tsplatz 5, 8010 Graz, Austria
\and Laboratoire d\'{}Astrophysique de Toulouse et Tarbes, UMR5572, CNRS et Universit{\'e} Paul Sabatier Toulouse 3, 57 avenue d\'{}Azereix, 65000 Tarbes France
\and Astronomical Institute of the Slovak Academy of Sciences, 05960 Tatransk{\'a} Lomnica, Slovakia
\and Institute of Mathematics, University of Vienna, Nordbergstra\ss e 15, 1090 Wien, Austria}

\date{Received 7 August 2009 /
Accepted 20 November 2009}
\abstract{Small-scale magnetic fields in the solar photosphere can be identified in high-resolution magnetograms or in the G-band as magnetic bright points (MBPs). Rapid motions of these fields can cause magneto-hydrodynamical waves and can also lead to nanoflares by magnetic field braiding and twisting. The MBP velocity distribution is a crucial parameter for estimating the amplitudes of those waves and the amount of energy they can contribute to coronal heating.}
{The velocity and lifetime distributions of MBPs are
derived from solar G-band images of a quiet sun region acquired by the Hinode/SOT instrument with different temporal and spatial sampling rates.} 
{We developed an automatic segmentation, identification and tracking algorithm to analyse G-Band image sequences to obtain the lifetime and velocity distributions of MBPs. The influence of temporal/spatial sampling rates on these distributions is studied and used to correct the obtained lifetimes and velocity distributions for these digitalisation effects.}
{After the correction of algorithm effects, we obtained a mean MBP lifetime of $(2.50 \pm 0.05)$ min and mean MBP velocities, depending on smoothing processes, in the range of (1~-~2)~$\mathrm{km~s^{-1}}$. Corrected for temporal sampling effects, we obtained for the effective velocity distribution a Rayleigh function with a coefficient of $(1.62 \pm 0.05)$ $\mathrm{km~s^{-1}}$. The $x$- and $y$- components of the velocity distributions are Gaussians. The lifetime distribution can be fitted by an exponential function.} {}


\maketitle

\section{Introduction}
Magnetic bright points (MBPs) are small-scale magnetic features, visible as bright
points in the solar photosphere. Located in intergranular
lanes, they are called ``bright points'' (BP), ``network bright points'' (NBP), ``G-band bright points''
when observed with a G-band filter, or filigree in a network region. MBPs are called ``flowers'' if they appear grouped in a roundly shaped formation \citep[see e.g.][]{2004A&A...428..613B}. The reported diameters of MBPs
range from 100 up to 300 km \citep[see e.g.][]{1983SoPh...87..243M,2004A&A...422L..63W,2009A&A...498..289U}, and the corresponding magnetic field strength reaches values of several kG \citep[as revealed by inversion techniques,][]{2007A&A...472..607B,2009ApJ...700L.145V}. MBPs appear at the merging
points of granules \citep{1992SoPh..141...27M} and display a
complex evolution. Pushed by granules, they are able to form groups, but merging and splitting of single features frequently occur as well \citep{1996ApJ...463..365B}.

It is well known that the MBP brightness in the G-band is in close relation with magnetic fields \citep{1992Natur.359..307K,1993SoPh..144....1Y,2001ApJ...553..449B,2004A&A...428..613B,2006ASPC..358...61B,2009ApJ...700L.145V}. This part of the electromagnetic spectrum comprises CH molecule lines centered around $\lambda$~=~430~nm. The increased brightness of magnetic active features in the
G-band is due to a decreased opacity \citep[see e.g.
][]{1999ASPC..184..181R,2001A&A...372L..13S,2003ApJ...597L.173S}. G-band images yield a better resolution than magnetograms, and thus provide good means to detect small features such as MBPs. However, one must be cautious, because also non-magnetic brightenings can be found \citep[e.~g. on top of granular fragments, see][]{2001ApJ...553..449B}.

MBPs are described either by dynamical or statical MHD theories. Early works on theoretical aspects were done e.g. by
\citet{1976SoPh...50..269S,1983ApJ...268..412O,1984A&A...139..426D}
and \citet{1984A&A...139..435D}. Simulations of dynamical processes in the photosphere were done by \citet[][]{1998ApJ...495..468S,2006ApJ...651..576S} and by \citet{2005ApJ...631.1270H}, who focused on dynamical chromospheric phenomena triggered by photospheric flux tubes.

MBPs are in many ways of interest. First of all it is supposed that rapid MBP footpoint motion can excite MHD waves. These waves could contribute in a significant way to the heating of the solar corona \citep[see e.g.][]{1993SoPh..143...49C,1994A&A...283..232M}. In addition to the wave heating processes also nanoflares can be triggered by these flux tube motions \citep[see e.g.][]{1983ApJ...264..642P,1988ApJ...330..474P}. Another research field in solar physics deals with the generation of the magnetic fields. Are small-scale fields like MBPs connected to the global magnetic field, or are they generated by small-scale magneto-convection \citep[see e.g.][]{2005A&A...429..335V}? Does the magnetic field interact on small scales with the solar granulation pattern and hence influence the solar irradiation and/or activity?

In this work we concentrate on the lifetime and velocity distributions of MBPs like several authors before: \citet{1983SoPh...85..113M,1994A&A...283..232M,1996ApJ...463..365B,2006SoPh..237...13M}. In addition to remeasuring these parameters with seeing-free observations from the 50~cm Hinode/SOT space-based telescope in the quiet sun, we investigate the dependence of the measurements on observational aspects like temporal and spatial sampling. This is important to understand why different studies yielded different results and how these differences could be explained. In Sect. 2 we give an overview of the data we used and on the stability of the Hinode satellite pointing. In Sect. 3 we describe the developed tracking algorithm. In Sect. 4 we present our results and the implications of different spatial/temporal sampling rates and a way to correct for digitalisation and algorithm effects. In Sect. 5 we discuss our results in the context of other studies, and in Sect. 6 we give a brief summary and our conclusions.

\section{Data}
We used two different data sets near the solar disc centre obtained by the solar optical telescope \citep[SOT; for a description see][]{2004SPIE.5487.1142I,2008SoPh..tmp...26S} on board
the Hinode satellite \citep[][]{2007SoPh..243....3K}. SOT has a
50 cm primary mirror limiting the spatial resolution by diffraction
to about 0.2 arcsec. Data set I, taken on March 10, 2007, has a field-of-view (FOV) of 55.8 arcsec by 111.6 arcsec with a spatial sampling of 0.108 arcsec per pixel. The complete time series covers a period of approximately 5 hours and 40 minutes, consisting of 645
G-band images recorded with a temporal resolution of about 32 seconds.
Data set II, from February 19, 2007, has a FOV of 27.7~arcsec by
27.7~arcsec with a spatial sampling of 0.054 arcsec per pixel and consists of 756 exposures over a period of roughly 2 hours and 20 minutes. Data set II has a better temporal resolution of about 11 seconds.
Both data sets cover a quiet sun FOV and were fully
calibrated and reduced by Hinode standard data reduction algorithms
distributed under solar software (SSW).

\begin{figure}
	\centering
		\includegraphics[scale=0.45]{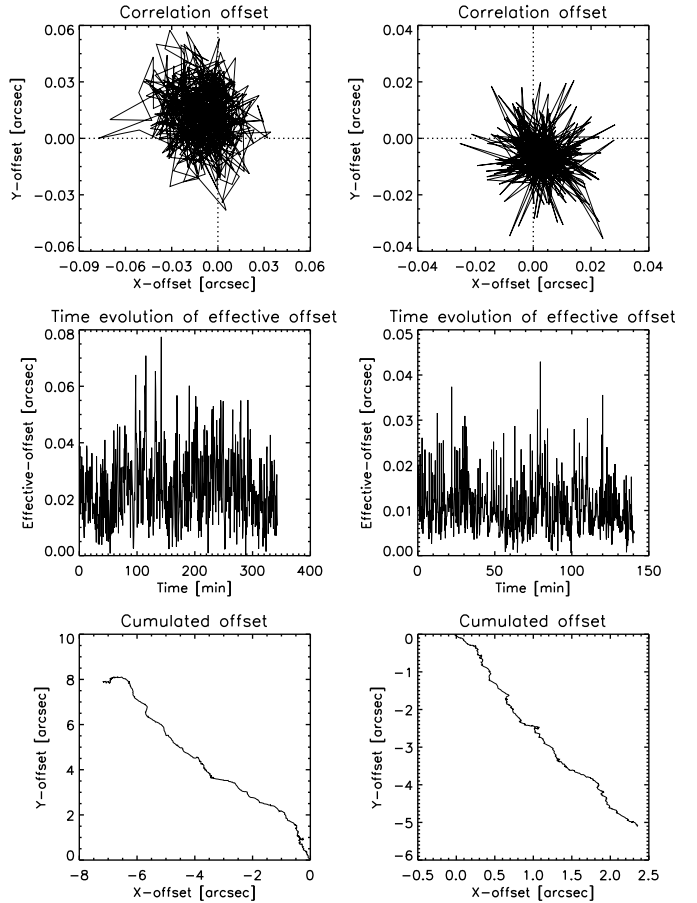}
		\caption{Hinode/SOT pointing stability for data sets I (left) and II (right). Top row: the displacement between an image and the succeeding one ($\Delta y$ versus $\Delta x$). The dotted lines indicate stable satellite pointing. It can be seen that the actual barycenter of the displacements is shifted diagonally out of this point (drift). Middle row: effective ($\sqrt{\Delta x^2+\Delta y^2}$) displacements versus time. Bottom row: cumulated displacements in $x$- and $y$-direction, which reflects the satellite pointing on the solar surface.}
	\label{figDisplacement}
\end{figure}

We focus now on the Hinode/SOT image stability. As MBP velocities are in the range of a few kilometers per second, the stability of the satellite pointing and the resulting image stability has to be checked very carefully. This was done by calculating the offsets between each image and the succeeding one by cross-correlation techniques (e.g. SSW routine \textit{cross\_corr}). Figure \ref{figDisplacement} illustrates the outcome. The calculated offsets between two frames were on average lower than one pixel ($0.2 \pm 0.1$ pixels). The offsets versus time is plotted in the second row of Fig. \ref{figDisplacement}. These agree with the SOT specified and measured image stability \citep[][]{2008ASPC..397....5I}. On the other hand, it is not only necessary that the images are coaligned from time step to time step, but there should also be no systematic drift of the satellite-pointing. In order to check this, we calculated the cumulated displacements. This showed that the satellite drifted away from the original pointing (see top and bottom row of Fig. \ref{figDisplacement}) with a mean speed of about 0.4 $\mathrm{km~s^{-1}}$, which lies in the range of MBP velocities. Therefore, we had to correct for this drift in the MBP data analysis.

\section{MBP tracking algorithm}
For the segmentation and identification of MBPs we used the algorithm described in \citet{2009A&A...498..289U}. This algorithm was further extended to track identified MBPs in subsequent images, as described below.
\begin{figure}
	\centering
	\includegraphics[width=.45\textwidth]{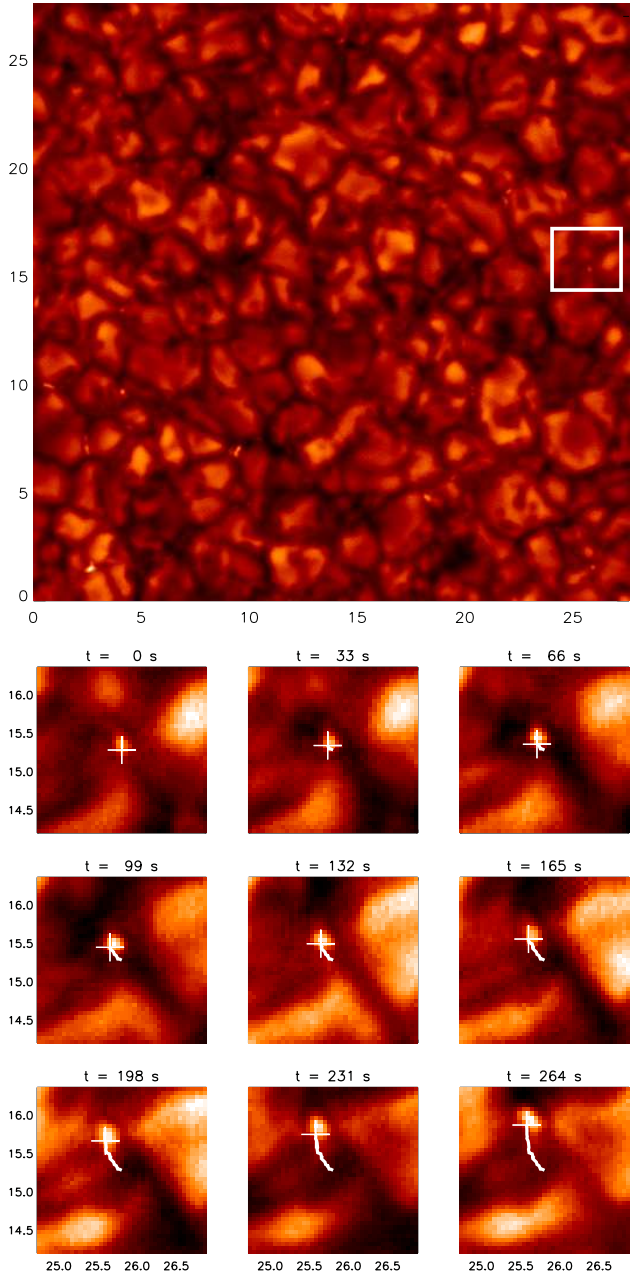}
		\caption{Top panel: Starting frame of one of the identified MBPs. Bottom panels: Evolution (time series) of the detected MBP in the rectangle. The white cross marks the derived brightness barycenter. The path of the MBP is shown as a white line. Positions are given in arcsec from the bottom left corner of the full frame.}
	\label{figsingle8}
\end{figure}

After the application of the segmentation and identification routines \citep{2009A&A...498..289U}, we obtain a set of single analysed images. Analysed images means that we have identified the MBP features and obtained relevant parameters like size, brightness and position. For dynamical parameters like lifetimes and velocities, it is necessary to analyse not only a single image but a series of images. The basic idea behind our tracking algorithm is simple. The algorithm takes the position of a detected MBP and compares it to the positions of MBPs in the succeeding image. If two MBPs match within a certain spatial range (which was set to 2.8 Mm for both data sets), the points are identified as the same MBP at different times. Therefore, a subsequent application of this comparison (for all images and all MBPs in the images) leads to complete time-series of MBP features. The algorithm is able to discern between isolated MBPs and groups of MBPs (this is explained in more detail in Sect. 4). The tracking algorithm has to consider several aspects as for instance noise and finite size of the images as well as the time span of the image sequence, to work correctly. All features which are identified either in a single image or in two images should be regarded as noise (or identification artefacts) and are therefore discarded \citep[for possible causes of noise features see][]{2009A&A...498..289U}. The spatial limitation of images can cause the time series of a MBP to be interrupted and splitted up. This occurs e.g. when a MBP moves across the border of the image. Thus, all MBPs that move too close to an image border are dismissed. The beginning and end of a time series should be treated the same way. As we do not know whether a MBP existed already before the first exposure was taken, or exists still after the last exposure was taken, all MBPs starting with the first image or ending with the last image are dismissed. All these effects have to be taken into account to avoid an underestimation of the obtained lifetimes. Figure \ref{figsingle8} shows an example of MBP tracking.

\section{Results}
\subsection{MBP lifetimes}
\begin{figure}
	\centering
		\includegraphics[scale=0.55]{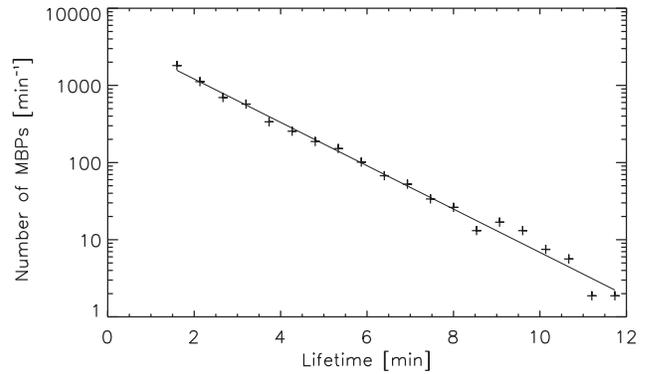}
		\caption{Histogram of the obtained MBP lifetimes of data set I together with an exponential fit (solid line). The histogram was normalised by dividing the counts in each bin by the binsize. The histogram is only shown up to the first zero crossing, i.e. there are a few longer-living MBPs than shown in the figure.}
	\label{figlifediff}
\end{figure}

\begin{figure}
	\centering
		\includegraphics[scale=0.55]{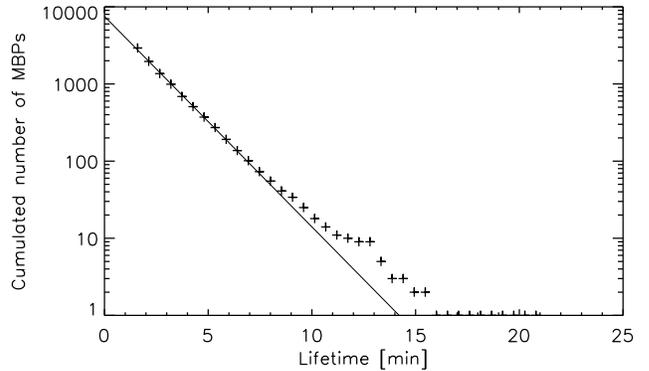}
		\caption{Cumulated number of MBPs versus lifetime fitted with an exponential function (solid line) for data set I. The values of the fit represent the number of MBPs that have a lifetime greater than or equal to the corresponding $x$-axis value $\left(\int^{\infty}_{t_x} \frac{dN}{dt}dt\right)$. Only the statistically robust bins ($>$ 100 data points) were considered for the fit.}
	\label{figlifecum}
\end{figure}

After identification and tracking of MBPs in the G-band image sequences, we derived the lifetime of each MBP. The lifetimes were measured by the time difference between the first and last detection of each MBP. For a correct interpretation, one should distinguish between isolated and grouped MBPs. An MBP that does not have any neighbours during its whole lifetime counts as an isolated point feature. A grouped MBP is a feature that has at least in one instance of its life another MBP in its vicinity. The vicinity was defined to be approximately 2800 km. In this paper, we concentrate on isolated MBPs, as it is quite difficult to obtain and interpret the lifetimes of grouped MBPs correctly. Grouped MBPs can split up and merge again. Therefore, many ``substrings'' in the group evolution can occur, and it is difficult to find a proper definition of the term ``lifetime''. Various kinds of definitions appear possible, e.g. using the average over all substrings, the substring with the longest duration, the time difference between first and last occurence of a substring. When two isolated features meet and form a group, the uniqueness of a substring is lost. If they split up again in the further evolution, it is unclear which one should be followed. In general the decision rules will thus strongly influence the derived lifetime distribution. Figure \ref{figlifediff} shows the measured lifetime histogram for data set I with normalised data bins. The number of MBPs with a certain lifetime as well as the cumulated (the total (integral) number $N(t)$ of all MBPs that have a lifetime of longer than $t$ minutes) number of MBPs (see Fig. \ref{figlifecum}) were fitted with an exponential function decreasing with lifetime (for an interpretation of the fit coefficients see Table \ref{explife}):
\begin{equation}\label{lifetimeequ}
    x(t)=a~\exp(-b\cdot t).
\end{equation}
\begin{table*}
\centering
\caption{Explanation of the different fitting parameters used for Eq. \ref{lifetimeequ}.}
\begin{tabular}{l l l l}
\hline
distribution type & fit parameter & representation & meaning \\
\hline
lifetimes&$x(t)$& $A(t)$& Number of MBPs with a lifetime of $t$ minutes in a 1 minute interval.\\
& & & Could be interpreted as ``activity'' (according to radio-activity):\\
& & & number of MBPs that decay within 1 minute at time $t$.\\
& $a$& $A(0)$& Starting activity (number of MBPs which will decay within the first min)\\
&$b$&$\lambda$&Decaying parameter (related to the mean lifetime $\tau=1/\lambda$)\\
\hline
cumulated& $x(t)$& $N(t)$& Number of MBPs with a lifetime greater than $t$\\
lifetimes& $a$& $N(0)$& Number of all MBPs\\
&$b$&$\lambda$&Decaying parameter\\
\hline
\end{tabular}
\label{explife}
\end{table*}
The exponentially decreasing number of MBPs can be caused by the physics of the process under estimation and/or by the used algorithm. If we take into account that the algorithm has a certain probability $p$ to find an MBP once, then the probability to find the same MBP in the next image again would be $p^2$. For $n$ images we achieve $p^n$ which leads to:
\begin{equation}\label{lifetimeequ1}
    p(t)=p^{\frac{t}{\Delta t}+1},
\end{equation}
where $t=n\cdot \Delta t$ is the time from the first MBP detection and $\Delta t$ is the time difference between two consecutive images. This can be easily transformed to:
\begin{equation}\label{lifetimeequ2}
    p(t)=p \cdot \exp\left(\frac{t}{\Delta t}~ \ln(p)\right).
\end{equation}

\begin{figure}
	\centering
		\includegraphics[scale=0.55]{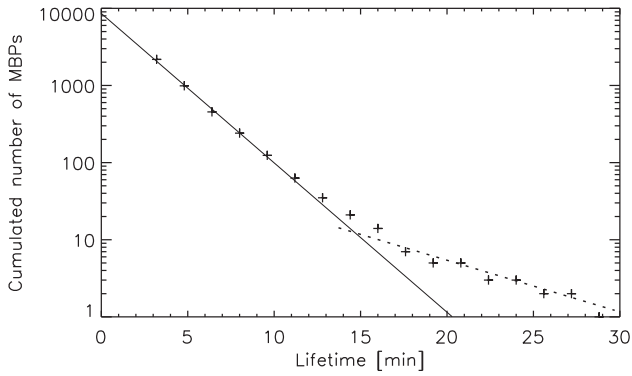}
		\caption{Same as Fig. \ref{figlifecum} but derived for a reduced sampling rate (only each third image was used). The corresponding temporal resolution is 96 seconds. The solid line shows a linear fit for the statistically more robust data. The dotted line shows a linear fit for a second ``unresolved'' distribution (for details see text).}
	\label{figlifecum3}
\end{figure}

\begin{figure}
	\centering
		\includegraphics[scale=0.55]{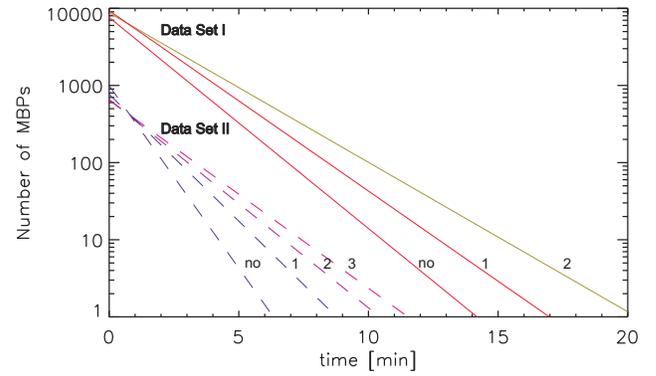}
		\caption{Cumulated number of MBPs versus time for the two different data sets and different temporal samplings. Fits belonging to data set I are shown by solid lines, data set II by dashed lines. The numbers indicate how many images were skipped to artificially decrease the temporal sampling.}
	\label{life}
\end{figure}

\begin{figure}
	\centering
		\includegraphics[scale=0.55]{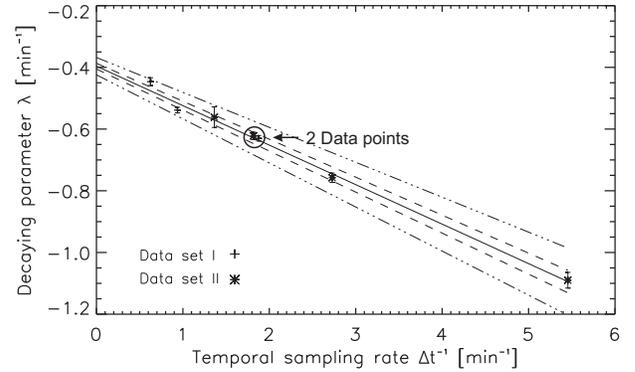}
		\caption{Decaying parameter $\lambda_{tot}$ derived from the fits plotted in Fig. \ref{life} as a function of the temporal sampling rate. A first order fit was applied, which yields the decaying rate $\lambda_p$ corrected for detection/algorithm effects (solid line). The derived physical decaying rate value is $\lambda_p~=~-0.40\pm0.01~\mathrm{min}^{-1}$, corresponding to a mean MBP lifetime $\tau~=~2.50\pm 0.05~\mathrm{min}$. Values obtained by data set I are shown as crosses, values derived by data set II are plotted by stars. Note that the two data points, derived by different data sets but with similar sampling rate, closely overlap. The 1-sigma bandwith of the fit is shown by dashed lines, the 3-sigma bandwith by dash/dotted lines.}
	\label{lifedecay}
\end{figure}

If we suppose that the real number of MBP versus lifetime distribution follows an exponential function, $N_0\exp(-\lambda\cdot t)$, we find for the measured distribution (for a more elaborate derivation of the measured distribution see the appendix):
\begin{eqnarray}
    N(t) & = & N_0 \cdot \exp(-\lambda_p~t)\cdot p(t) \\
    & = & N_0\cdot p \cdot \exp(-\lambda_p~t)\cdot \exp\left(\frac{t}{\Delta t}~ \ln(p)\right)\\
    & = & N_0 \cdot p \cdot \exp\left(\left(-\lambda_p+\frac{1}{\Delta t}~ \ln(p)\right)~t\right)\\
    & = & N_0 \cdot p \cdot \exp(-\lambda_{tot}~t).
\end{eqnarray}
In this context $N(t)$ denotes the measured number of MBPs with a lifetime larger than $t$. $N_0$ stands for a ``fictitious'' starting number of MBPs. MBPs are formed and decay constantly. If we could force all MBPs to be formed in the first exposure and then could turn off the forming process, $N_0$ would denote the number of features to be found in the first image. $\lambda_p$ gives us the physical decaying parameter, whereas $\lambda_{tot}$ is the ``total'' decaying parameter derived from the measurements with the tracking algorithm. A comparision of Eqs. (7) and (6) gives the relationship between the measured decaying constant ($\lambda_{tot}$) and the temporal sampling rate $\Delta t$ as:
\begin{equation}
\lambda_{tot}=\lambda_p-ln(p)~\frac{1}{\Delta t}.
\end{equation}
If we obtain the total decaying constant for data sets with a different temporal sampling $\Delta t$, we should be able to recover the true physical decaying rate ($\lambda_p$) as well as the detection probability of our algorithm ($p$).

\begin{figure*}
	\centering
		\includegraphics[scale=1.1]{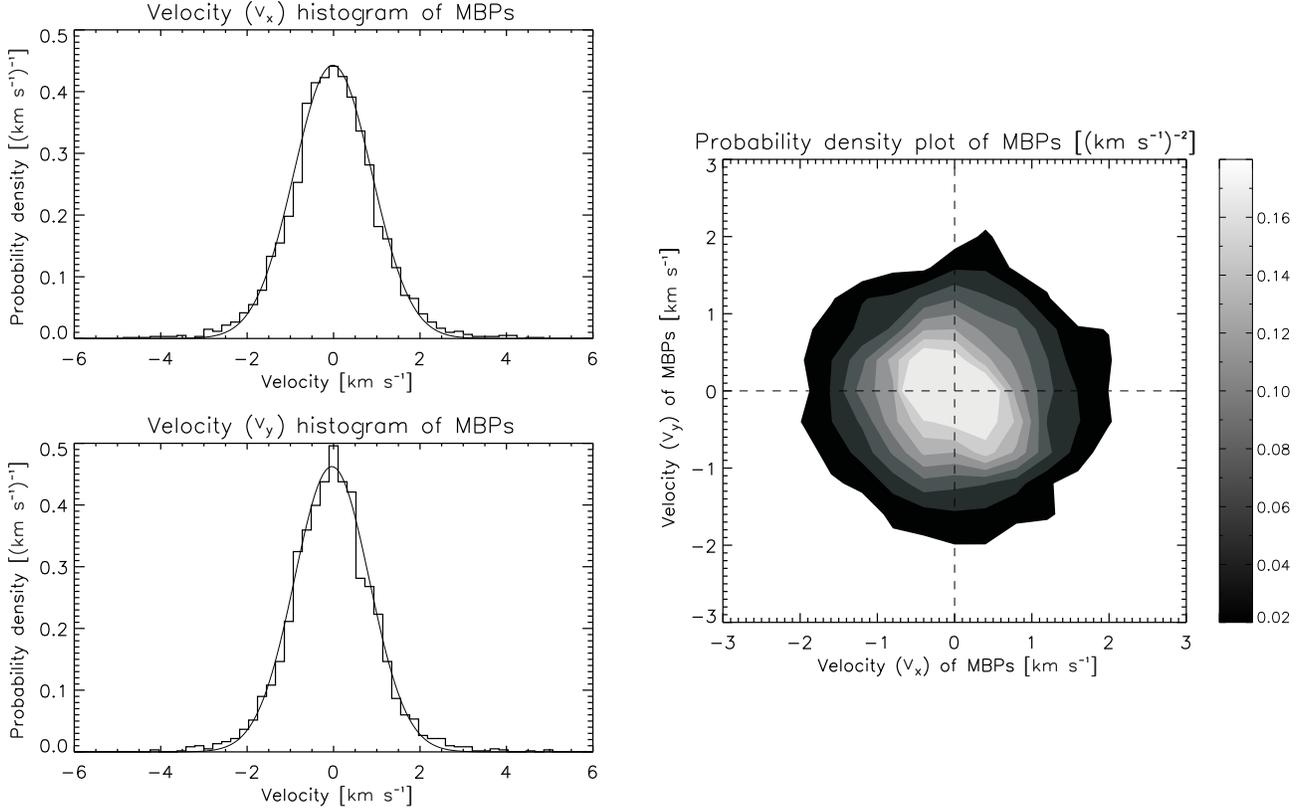}
		\caption{The top left panel shows the $x$-component of the velocity of identified single MBPs together with a Gaussian fit for data set I. The fit parameters are: $\mu~=~-0.02~\mathrm{km~s}^{-1}$ and $\sigma~=~0.89~\mathrm{km~s}^{-1}$. This distribution was derived by skipping two images (giving a temporal resolution of 96 seconds). The $y$-component is shown below. The fit parameters are: $\mu~=~-0.04~\mathrm{km~s}^{-1}$ and $\sigma~=~0.85~\mathrm{km~s}^{-1}$. The probability density is displayed on the right side. It can be seen that there is no preferential direction for the MBPs movement.}
	\label{figvelall}
\end{figure*}
To test wheter this relationship really holds, we artificially reduced the temporal sampling of our data sets. For each of these realisations, we got a distribution similar to Fig. \ref{figlifecum}.
 We artificially decreased data set I by interleaving one or two images (corresponds to images every 64 and 96 seconds, respectively) and data set II by interleaving one, two and three images (corresponds to images every 22, 33 and 44 seconds, respectively). Interestingly on thus other realisations is that a tail is forming out of the measurements cloud of the long-living MBPs of Fig. \ref{figlifecum}. This can be seen e.g. in Fig. \ref{figlifecum3}, which shows the cumulated lifetime for an image cadence of 96 seconds. This could represent a second (unresolved) distribution, indicated by the dashed line. Since we only found about 20 MBPs with lifetimes longer than 15 minutes, we did not investigate this tail further. The resulting exponential fits to the histograms are shown together with the original ones in Fig. \ref{life}. It can be seen that the slope of the distributions is very similar for skipping 0 images in data set I to skipping 2 images in data set II, which relate both to a similar temporal sampling of about 30 seconds (32 to 33 seconds; compare also Fig. \ref{lifedecay}). The obtained fit parameters are listed in Table~\ref{tablelifetimes}. Decreasing the temporal sampling leads to a flattening of the exponential slope. In Fig. \ref{lifedecay} we plot the derived decaying constants $\lambda_{tot}$ versus the temporal resolution, which follows a straight line. From the linear fit to these curve we derived $\lambda_p=-0.40 \pm 0.01 \mathrm{min^{-1}}$, which corresponds to a mean MBP lifetime of $2.5 \pm 0.05$ min. The second fit parameter gives us after some transformations the identification probability $p$ of the MBP features in single images. This probability has a value of about 90\%.

\begin{table}
  \centering
  \caption{Overview of the measured lifetimes of MBPs.}\label{lifetime_table}
  \begin{tabular}{l l l}
    \hline
    Temporal cadence & Fit parameter & Fit parameter   \\
     $\Delta t$ [$\mathrm{s}$]  & $N_0$ & $\lambda$ [$\mathrm{min^{-1}}$]  \\
    \hline
    11 &$ 990 \pm 70 $&  $-1.09 \pm 0.03$  \\
    22 &$ 770 \pm 40 $&  $-0.76 \pm 0.02$ \\
    33 &$ 660 \pm 30 $&  $-0.62 \pm 0.01$ \\
    44 &$ 640 \pm 90 $& $-0.56 \pm 0.03$ \\
    \hline
    32 &$ 7 600 \pm 250$ &$ -0.63 \pm 0.01$\\
    64 &$ 9 300 \pm 450$ & $-0.54 \pm 0.01$ \\
    96 &$ 8 700 \pm 750$ & $-0.45 \pm 0.01$ \\
    \hline
  \end{tabular}
  \label{tablelifetimes}
  \end{table}

\subsection{Velocities}
\begin{figure}
	\centering
		\includegraphics[scale=0.55]{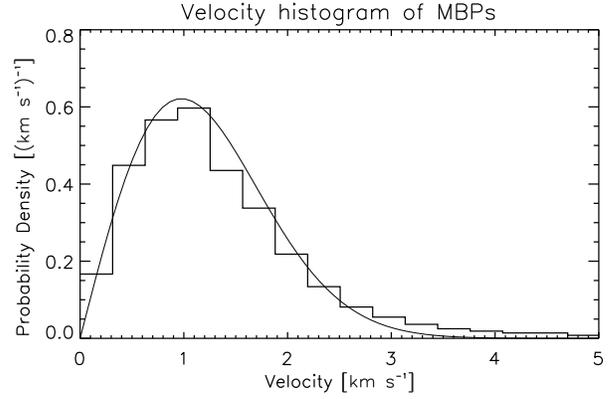}
		\caption{Effective velocity $v=\sqrt{{v_x}^2+{v_y}^2}$ of the MBPs of data set I is shown together with a Rayleigh fit. The statistic is defined by skipping two images between measurements. The fit parameter $\sigma$ is in this case 0.88 $\mathrm{km~s^{-1}}$. The fit coefficients for different smoothing levels are summarised in Table \ref{velocity108}.}
	\label{figrayleigh}
\end{figure}

The MBP velocity distributions were derived by measuring the movement of the MBPs brightness barycenter.
The barycenter positions were corrected for the satellite pointing drift as discussed in Sect. 2. We smoothed the velocities by reducing the available image cadence. The velocity distributions can be fitted by Gaussian distributions for the $x$- and $y$-velocity components (see Fig. \ref{figvelall}), respectively. The effective velocity ($v=\sqrt{{v_x}^2 + {v_y}^2}$) was fitted by a Rayleigh distribution (see Fig. \ref{figrayleigh}):
\begin{equation}\label{rayleigh_equ}
    f(v,\sigma)=\frac{v}{\sigma^2}\exp\left(-\frac{v^2}{2\sigma^2}\right).
\end{equation}
In this notation, $\sigma$ is the standard deviation of the distribution. 

\begin{table}
  \centering
  \caption{Overview of the velocity fit coefficients for the Rayleigh distributions (shown in Figs. \ref{figvel_smooth} and \ref{figveldrift}) for both data sets.}
  \begin{tabular}{l l l l}
    \hline
    Spat. sampling & Temp. cadence  & Fit parameter & Mean value \\
    ~[$\mathrm{arcsec/pixel}$]  &$\Delta t$ [$s$] &$\sigma$ [$\mathrm{km~s^{-1}}$]& $\mu$ [$\mathrm{km~s^{-1}}$] \\
    \hline
    0.108 & 32 & 1.12&  $1.59\pm 1.20$   \\
     & 64 & 0.95&  $1.32\pm 0.91$   \\
     & 96 & 0.88&  $1.17\pm 0.76$  \\
     & 128 & 0.81&  $1.08\pm 0.67$   \\
    & 160 & 0.78&  $1.03\pm 0.61$   \\
     & 192 & 0.72&  $0.94\pm 0.56$   \\
    \hline
    0.054 & 11 & 1.36&  $1.77\pm 1.22$   \\
     & 22 & 1.14&  $1.49\pm 0.90$   \\
     & 33 & 1.04&  $1.35\pm 0.84$  \\
     & 44 & 1.01&  $1.29\pm 0.77$   \\
     & 55 & 0.98&  $1.25\pm 0.73$   \\
     & 66 & 0.94&  $1.21\pm 0.70$   \\
    \hline
  \end{tabular}
  \label{velocity108}
\end{table}

Figure \ref{figvel_smooth} shows the velocity distributions and Rayleigh fits for data set I obtained by different temporal sampling rates. For detailed information about the fit coefficients derived for different smoothing levels we refer to Table \ref{velocity108}. Figure \ref{figveldrift} shows the Rayleigh fits of these distributions all together in one plot. It can be seen in both figures that the velocity distribution shows a scaling behaviour with the $\sigma$-parameter. As we have only changed the temporal sampling rates, we know that there has to be some relation between these parameters. We have tried to find this relation between the $\sigma$ coefficient of the Rayleigh distribution and the corresponding temporal ($\Delta t$) sampling rate. We applied the following empirical relationship:
\begin{equation}\label{velsmoothequ}
    \sigma^{-1}={\sigma_0}^{-1}+\frac{\sqrt{\Delta t}}{\sqrt{D}}.
\end{equation}
In units of measure $D$ corresponds to a diffusion parameter of an ensemble of MBPs. This means that this parameter should describe the random movement and spreading of a population of MBPs (not to be mixed up with the diffusion of the magnetic field of a single MBP). The application of this relation as a fit function on our data sets is shown in Fig. \ref{nuplot}. We obtained the fit parameters $\sigma_0=(1.62 \pm 0.05$) $\mathrm{km~s^{-1}}$ and $D=(350 \pm 20$) $\mathrm{km}^2~s^{-1}$. $D$ agrees within the errors with the diffusion coefficient in \citet{1998ApJ...506..439B}, who reported a value of 285~$\mathrm{km^2~s^{-1}}$ for a quiet Sun FOV. The interesting question is which effective velocities would be measured, if we could continously observe the Sun rather than just observing the positions of the MBPs at certain instances of time? Therefore, we can calculate the limit for ${\Delta t \rightarrow0}$ that will result in $\sigma_0$. $\sigma_0$ should therefore be the ``true'' standard deviation for the velocity distribution independent of temporal sampling. Interestingly this value is more than twice (2.2) as large as the smallest measured sigma value and still 1.4 times as large as the highest measured value (without consideration of the sigma value for the highest cadence as this value seems to be an outlier), see also Table \ref{velocity108}. From this extrapolated ``true'' sigma parameter for the Rayleigh distribution we can easily derive a new ``true'' mean velocity by the equation for the theoretical mean of this distribution ($
	\mu=\sigma\cdot\sqrt{(\pi/2)}$), giving a value of 2.0 $\mathrm{km~s^{-1}}$. Figure~\ref{reconst} shows the Rayleigh distribution created with the estimated true $\sigma$-parameter.

\section{Summary and Discussion}
\begin{figure*}
	\centering
		\includegraphics[scale=1]{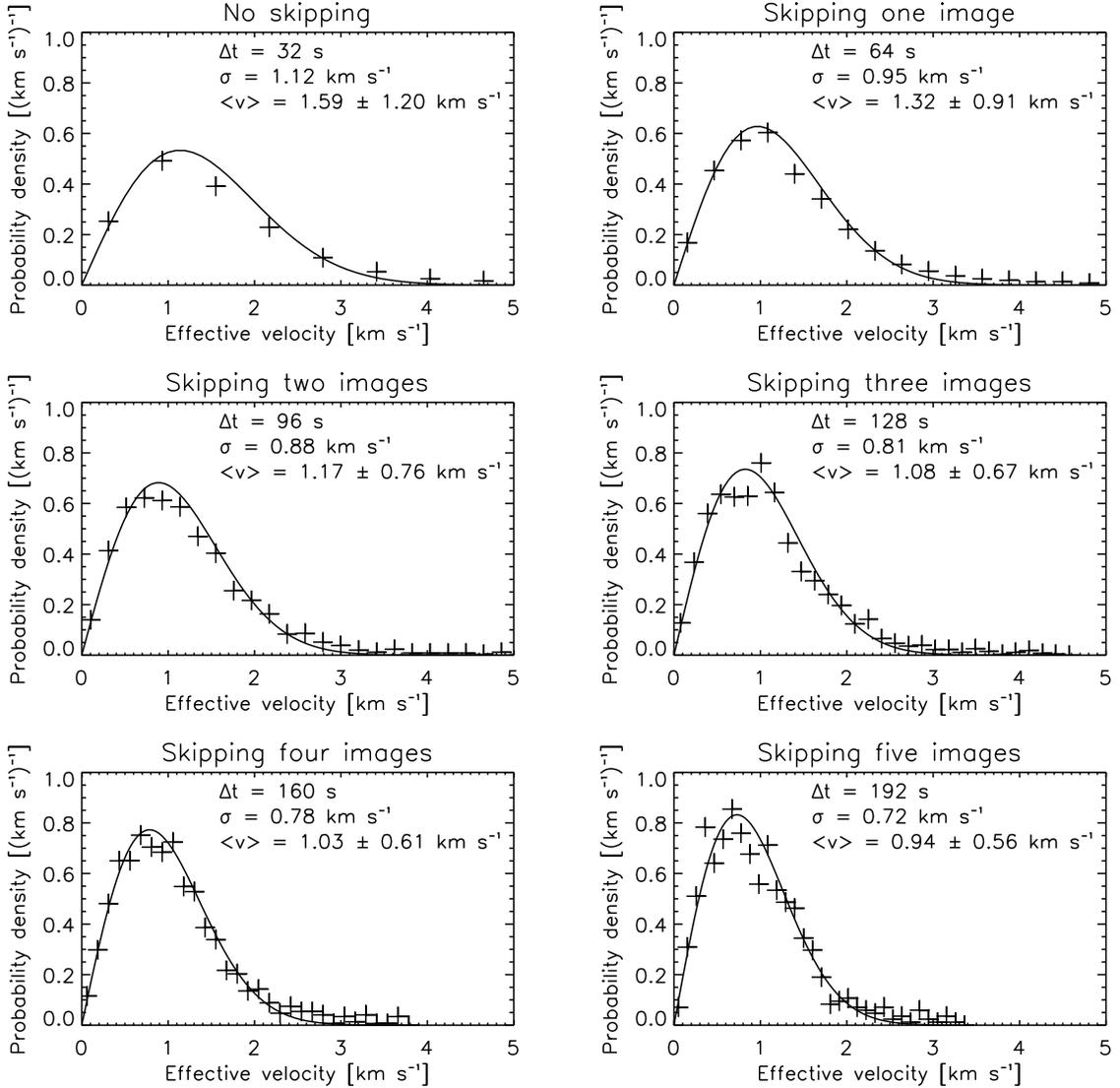}
		\caption{MBP velocity distribution for different smoothing levels for data set I. The temporal resolution is reduced from the top left panel (each image used; $\Delta t=32~\mathrm{s}$) to the bottom right (five images are skipped between successive frames; $\Delta t=192~\mathrm{s}$). Solid curves represent the fits performed by Rayleigh functions. The fit parameters are given in Table \ref{velocity108}.}
	\label{figvel_smooth}
\end{figure*}
We have further extended the detection and identification algorithm for MBPs described in \citet{2009A&A...498..289U}. It consists of an additional processing step, which enables us to analyse the time series of G-band images. We found MBP velocity and lifetime distributions in accordance with earlier publications. The lifetime distribution shows an exponential decrease for longer lifetimes. The decaying parameter was found to be 0.4 $\mathrm{min}^{-1}$ corresponding to a mean lifetime of 2.5 min. We outlined a way to estimate the identification probability of the algorithm (about 90\%) and to correct the lifetimes for the effects of imperfect detection.

If we compare our lifetime-result with previous studies (see Table \ref{tablelifetimes2}), we see that lifetimes obtained by automated feature tracking algorithms tend to be shorter than those measured by visual identification and tracking. The reason for this could be that if we identify a feature in a certain exposure visually, we would try to find the same feature in the previous and subsequent images. Since we presume that in the next or previous image the same MBP feature will be found, we tend to identify structures as beginning or decaying MBPs (even when they are too weak to be identified by an automated identification routine or are possiblly unrelated features) that we would not identify if we had no clue about the temporal evolution. The results also depend on how the network and inter-network fields are defined and observed. In \citet{2008ApJ...684.1469D} small-scale fields were first identified in magnetograms; on the other hand, in \citet{2005AA...441.1183D} MBPs were used as proxies for the magnetic field resulting in a difference of the derived lifetimes of a factor of three.

  \begin{table}
  \centering
  \caption{Overview on MBP lifetimes in earlier studies. In \citet{2008ApJ...684.1469D} magnetograms were used for the identification of the small-scale magnetic fields (instead of MBPs).}
 \begin{tabular}{l l}
   \hline
  Paper &  reported MBP lifetime [min]\\
    \hline
    \citet{2008ApJ...684.1469D}& 10\\
 \citet{2006SoPh..237...13M} &  4.4 ($\pm 2.4$)\\
 \citet{2005AA...441.1183D}& 3.5\\
  \citet{2004ApJ...609L..91S}&  $<$ 10\\
   \citet{1996ApJ...463..365B} &  $\sim(6-8)$\\
   this study & 2.5\\
  \hline
  \end{tabular}
  \label{tablelifetimes2}
\end{table}

Our results suggest that the different values for MBP lifetimes are related to different data sets (sampling) and methods but may also depend on the amount of magnetic flux in the region of interest, i.e. depending on the local physical properties of the solar photosphere. Near large magnetic flux regions such as plages or sunspots, MBP lifetime could be larger as a result than in quiet Sun regions. This can be explained if one keeps in mind that the dissipation of large amounts of flux would in turn also take longer. Our results, obtained from the analysis of isolated MBPs agree with \citet{2005AA...441.1183D}. In this work, the authors excluded network regions (i.e., MBP chains) from the analysis through the use of Ca II H maps. The selection of isolated bright features in both the \citet{2005AA...441.1183D} analysis and the analysis reported here is the most plausible argument to explain the agreement above cited. The second unresolved distribution at larger lifetimes in Fig.5 may be related to small patches of active network in our datasets. Although we aimed to exclude active patches (characterised by grouped MBPs) in our study, it could happen that some points were still identified (we note that Fig. \ref{figlifecum3} shows in total only about 20 of these events). The derived effective velocity distribution was fitted by a Rayleigh distribution. We saw that the velocity distribution depends on the temporal resolution. Depending on this sampling effect we estimated the Rayleigh fit parameter to be in the range of 0.7 to 1.4 $\mathrm{km~s^{-1}}$. This corresponds to mean velocities of 0.9 to 1.8 $\mathrm{km~s^{-1}}$. The true $\sigma$-parameter for the Rayleigh velocity distribution was estimated to be 1.6 $\mathrm{km~s^{-1}}$, which corresponds to a mean velocity of about 2.0 $\mathrm{km~s^{-1}}$.

To estimate the velocity of the MBPs we used the brightness barycenter to get their positions and subsequently their velocities. Alternatively we could have measured the movement of the brightest pixel of an identified feature. Both methods have their advantages and disadvantages. Measuring the movement by tracking the brightest pixel of a feature is unique and not dependant on the shape of a feature. On the other hand this method leaves only a few possibilities of movement due to the achievable spatial sampling (in principle: up/down, left/right, and diagonal; one or two pixels). Therefore, it would be hard to gain a ``real'' velocity distribution. By taking the method of tracking the barycenter of brightness, many more possibilities are available, as the barycenter coordinates can be determined on sub-pixel resolution. However, this method is sensitive to the morphology of the feature and therefore on the definition of the size of the feature \citep[see also][on different possibilities of size definition]{2009Hvar}. 

Our tracking algorithm estimates the path of a MBP by comparing the positions of a MBP in two consecutive images. Therefore one is obliged to choose a certain comparison range, which could be easily estimated by multiplying the maximum anticipated velocity (about 4 $\mathrm{km~s^{-1}}$) by the temporal resolution. If the range is set too small, time series of MBPs will be broken up, causing the lifetimes and velocities to be underestimated. On the other hand, by choosing the range to large, unrelated features will be connected to a time series, which leads to an overestimation of lifetimes and velocities.

To get a real velocity distribution rather than just a few velocity ''states''\footnote{A state is a possible velocity value. As we are working with discretised data, there are no arbritrary displacement possibilities and therefore also the velocity values are not arbritary. Some velocity values are a lot more favourable (no movement, one pixel in a direction,\dots) than others. Calculating displacements by barycenters instead of calculating them by the difference of the brightest pixel helps, but as the sizes are very small the displacement states mentioned before are still more favourable.} we have to smooth our obtained velocity values. The smoothing of the distributions has several effects. First of all the distribution gets ``closer'' to the real distribution as there are now more possible velocity states. On the other hand the measured velocity decreases and the distribution gets narrower. This can be explained by the movement of a feature on the solar disc. This movement consists of two parts, a deterministic movement and a chaotic movement component. The deterministic movement is due to large scale flows, e.g., supergranular flows, meridional circulation etc.. These can cause MBPs to move over larger spatial scales and longer time scales with preferred directions and velocities than just moving randomly. The chaotic movement is due to random processes and corresponds to a ``zigzag'' path of a feature on the solar disc \citep[for implications of this random walk on the diffusive magnetic field transport see e.g.][]{1998ApJ...506..439B,2007Ap&SS.312..343S,2009Ap&SS.tmp..104S}. This yields a dependence of the length of the path on the temporal resolution of the observations. Consequently the shorter the temporal lags between the observations, the more accurate is the measurement of the path and the longer the path  will be (one could observe more zigzag steps).
So we have learned that by smoothing (artificially decreasing the time cadence), the derived MBP velocity decreases. This is not only true for MBP movements but holds also for other motions, as can be seen e.g. \citet{2009A&A...493L..13A} who investigated granular flow field maps and found a similar behaviour. On the other hand, if we do not smooth, we probably get velocities which are too high. This is due to the effect that the temporal to spatial sampling has to have a certain ratio. If we have a temporal sampling which is too high with respect to the spatial sampling (for a given velocity), the MBP would stand still for several exposures. If the feature finally moves, it jumps from one pixel to another and the derived velocity will be too high, since in reality the feature moves more or less continuously all the time. On the other hand, if the time resolution is too low (smoothing too high), we would loose much of the chaotic movement and derive a ``mean'' velocity which is too low. Figure \ref{figvel_smooth} shows this effect. We see that in the first cases the distribution is broader with a high velocity tail. The last two plots show that the obtained measurement points now concentrate more or less at average velocities (the distribution gets significantly narrower) and only a small number of high velocities can be found.  

\begin{figure}
	\centering
		\includegraphics[scale=0.55]{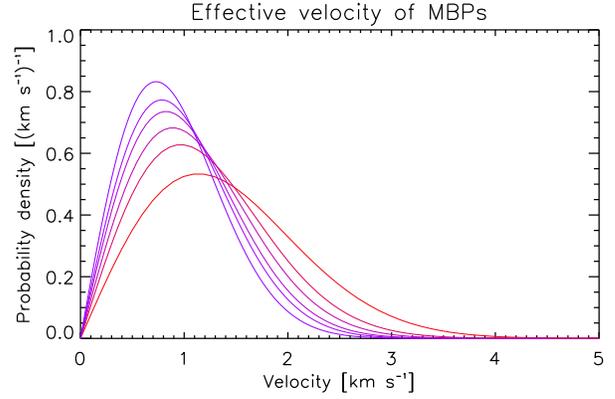}
		\caption{Comparison of all Rayleigh function fits to the velocity distributions plotted in Fig. \ref{figvel_smooth}. The broadness of the distribution as well as the mean velocity of the MBPs decreases with increasing smoothing level (from red to blue), i.e. lower time cadence.}
	\label{figveldrift}
\end{figure}

\begin{figure}
	\centering
		\includegraphics[scale=0.55]{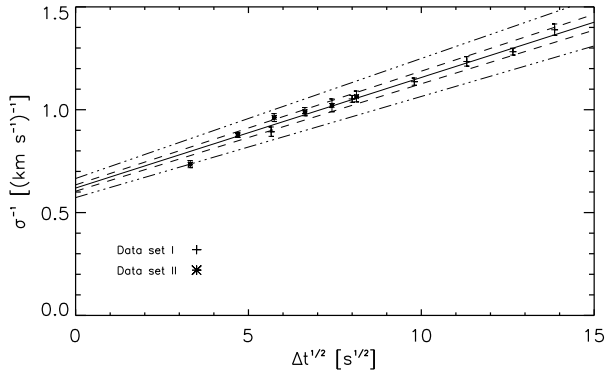}
		\caption{The measured Rayleigh $\sigma$-fit parameters are plotted versus the square root of the temporal sampling rate together with a first order fit. The dashed lines represent the 1-sigma bandwidth of the resulting fit. The dashed-dotted lines give the 3-sigma bandwidth. We interpret the slope of the fit to correspond to the diffusion process the MBPs are undergoing, whereas the other fit constant ($\sigma_0$) corresponds to the ``true'' Rayleigh distribution.}
	\label{nuplot}
\end{figure}

\begin{figure}
	\centering
		\includegraphics[scale=0.55]{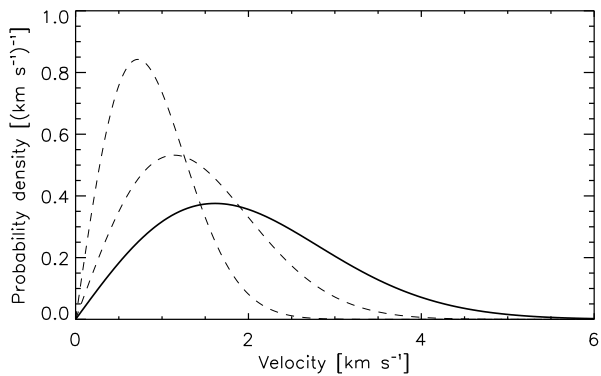}
		\caption{The obtained Rayleigh distribution for the estimated true $\sigma$ parameter (solid line). It can be seen that in this distribution large velocities are more feasible than in the original measured distributions (see also Figs. \ref{figvel_smooth} and \ref{figveldrift}). The original (measured) distributions span between the two distributions which are indicated by the dashed line (without consideration of the outlier $\sigma=1.36$ $\mathrm{km~s^{-1}}$ of Fig. \ref{nuplot}).}
	\label{reconst}
\end{figure}

Another interesting aspect is that not only the shape of the distribution but also the mean value of the distribution is changed. This is due to the fact that we have a non-symmetric function. In the process of smoothing, more of the high values are redistributed to lower values than vice versa. This is the mathematical explanation. The technical interpretation is that we loose a part of the MBPs path by reducing the temporal sampling. This leads to the shift of the distribution which can be seen in Fig. \ref{figveldrift}.

\section{Conclusion}
In this study we found MBP lifetimes (corrected mean value $2.50 \pm 0.05$ min) and velocities (corrected Rayleigh fit parameter $1.62 \pm 0.05$ $\mathrm{km~s^{-1}}$) in the range of earlier reported values. Additionally we investigated the relationship between image cadence and obtained lifetimes and velocities. Using the derived relations we were able for the first time to correct the measured velocity distribution for this influence. The corrected velocity distribution shows many more fast moving MBPs (v $>$ 4 $\mathrm{km s}^{-1}$) than the original measured distribution (12\% for the corrected distribution; 0.2\% for the measured distribution with the highest $\sigma$; see also Fig. \ref{reconst}). The obtained fraction of fast moving MBPs could play a crucial role in AC coronal heating models. In fact, fast moving MBPs have a large impact on the amount of energy available for AC heating processes. This was outlined in the theoretical work of \cite{1993SoPh..143...49C}; here the authors individuated in `fast motions' ($\simeq3~\mathrm{km s}^{-1}$) of magnetic footpoints a potential source of energy to support the coronal heating.

\begin{acknowledgements}
We are grateful to the Hinode team for the possibility to use their data. Hinode is a Japanese mission developed and launched by ISAS/JAXA, collaborating with NAOJ as a domestic partner, NASA and STFC (UK) as international partners. Scientific operation of the Hinode mission is conducted by the Hinode science team organized at ISAS/JAXA. This team mainly consists of scientists from institutes in the partner countries. Support for the post-launch operation is provided by JAXA and NAOJ (Japan), STFC (U.K.), NASA (U.S.A.), ESA, and NSC (Norway).
This work was supported by FWF \emph{Fonds zur F{\"o}rderung wissenschaftlicher Forschung} grant P17024.
D. U. and A. H. are grateful to the {\"O}AD \emph{{\"O}sterreichischer Austauschdienst} for financing research visits at the Pic du Midi Observatory. M.R. is grateful to the Minist{\`e}re des Affaires Etrang{\`e}res et Europ{\'e}ennes,
for financing a research visit at the University of Graz.
This work was partly supported by the Slovak Research and
Development Agency SRDA project APVV-0066-06 (JR).
We are grateful to the anonymous referee for her/his very detailed remarks and comments, which helped us to present the outcome of this work in a clearer way.

\end{acknowledgements}
\bibliographystyle{aa}

\begin{thebibliography}{41}
\expandafter\ifx\csname natexlab\endcsname\relax\def\natexlab#1{#1}\fi

\bibitem[{{Attie} {et~al.}(2009){Attie}, {Innes}, \&
  {Potts}}]{2009A&A...493L..13A}
{Attie}, R., {Innes}, D.~E., \& {Potts}, H.~E. 2009, \aap, 493, L13

\bibitem[{{Beck} {et~al.}(2007){Beck}, {Bellot Rubio}, {Schlichenmaier}, \&
  {S{\"u}tterlin}}]{2007A&A...472..607B}
{Beck}, C., {Bellot Rubio}, L.~R., {Schlichenmaier}, R., \& {S{\"u}tterlin}, P.
  2007, \aap, 472, 607

\bibitem[{{Berger} {et~al.}(1998){Berger}, {L{\"o}fdahl}, {Shine}, \&
  {Title}}]{1998ApJ...506..439B}
{Berger}, T.~E., {L{\"o}fdahl}, M.~G., {Shine}, R.~A., \& {Title}, A.~M. 1998,
  \apj, 506, 439

\bibitem[{{Berger} {et~al.}(2004){Berger}, {Rouppe van der Voort},
  {L{\"o}fdahl}, {Carlsson}, {Fossum}, {Hansteen}, {Marthinussen}, {Title}, \&
  {Scharmer}}]{2004A&A...428..613B}
{Berger}, T.~E., {Rouppe van der Voort}, L.~H.~M., {L{\"o}fdahl}, M.~G.,
  {et~al.} 2004, \aap, 428, 613

\bibitem[{{Berger} \& {Title}(1996)}]{1996ApJ...463..365B}
{Berger}, T.~E. \& {Title}, A.~M. 1996, \apj, 463, 365

\bibitem[{{Berger} \& {Title}(2001)}]{2001ApJ...553..449B}
{Berger}, T.~E. \& {Title}, A.~M. 2001, \apj, 553, 449

\bibitem[{{Bharti} {et~al.}(2006){Bharti}, {Jain}, {Joshi}, \&
  {Jaaffrey}}]{2006ASPC..358...61B}
{Bharti}, L., {Jain}, R., {Joshi}, C., \& {Jaaffrey}, S.~N.~A. 2006, in
  Astronomical Society of the Pacific Conference Series, Vol. 358, Astronomical
  Society of the Pacific Conference Series, ed. R.~{Casini} \& B.~W. {Lites},
  61

\bibitem[{{Choudhuri} {et~al.}(1993){Choudhuri}, {Auffret}, \&
  {Priest}}]{1993SoPh..143...49C}
{Choudhuri}, A.~R., {Auffret}, H., \& {Priest}, E.~R. 1993, \solphys, 143, 49

\bibitem[{{de Wijn} {et~al.}(2008){de Wijn}, {Lites}, {Berger}, {Frank},
  {Tarbell}, \& {Ishikawa}}]{2008ApJ...684.1469D}
{de Wijn}, A.~G., {Lites}, B.~W., {Berger}, T.~E., {et~al.} 2008, \apj, 684,
  1469

\bibitem[{{de Wijn} {et~al.}(2005){de Wijn}, {Rutten}, {Haverkamp}, \&
  {S{\"u}tterlin}}]{2005AA...441.1183D}
{de Wijn}, A.~G., {Rutten}, R.~J., {Haverkamp}, E.~M.~W.~P., \&
  {S{\"u}tterlin}, P. 2005, \aap, 441, 1183

\bibitem[{{Deinzer} {et~al.}(1984{\natexlab{a}}){Deinzer}, {Hensler},
  {Schuessler}, \& {Weisshaar}}]{1984A&A...139..426D}
{Deinzer}, W., {Hensler}, G., {Schuessler}, M., \& {Weisshaar}, E.
  1984{\natexlab{a}}, \aap, 139, 426

\bibitem[{{Deinzer} {et~al.}(1984{\natexlab{b}}){Deinzer}, {Hensler},
  {Schussler}, \& {Weisshaar}}]{1984A&A...139..435D}
{Deinzer}, W., {Hensler}, G., {Schussler}, M., \& {Weisshaar}, E.
  1984{\natexlab{b}}, \aap, 139, 435

\bibitem[{{Hasan} {et~al.}(2005){Hasan}, {van Ballegooijen}, {Kalkofen}, \&
  {Steiner}}]{2005ApJ...631.1270H}
{Hasan}, S.~S., {van Ballegooijen}, A.~A., {Kalkofen}, W., \& {Steiner}, O.
  2005, \apj, 631, 1270

\bibitem[{{Ichimoto} {et~al.}(2008){Ichimoto}, {Katsukawa}, {Tarbell}, {Shine},
  {Hoffmann}, {Berger}, {Cruz}, {Suematsu}, {Tsuneta}, {Shimizu}, \&
  {Lites}}]{2008ASPC..397....5I}
{Ichimoto}, K., {Katsukawa}, Y., {Tarbell}, T., {et~al.} 2008, in Astronomical
  Society of the Pacific Conference Series, Vol. 397, First Results From
  Hinode, ed. S.~A. {Matthews}, J.~M. {Davis}, \& L.~K. {Harra}, 5--+

\bibitem[{{Ichimoto} {et~al.}(2004){Ichimoto}, {Tsuneta}, {Suematsu},
  {Shimizu}, {Otsubo}, {Kato}, {Noguchi}, {Nakagiri}, {Tamura}, {Katsukawa},
  {Kubo}, {Sakamoto}, {Hara}, {Minesugi}, {Ohnishi}, {Saito}, {Kawaguchi},
  {Matsushita}, {Nakaoji}, {Nagae}, {Sakamoto}, {Hasuyama}, {Mikami},
  {Miyawaki}, {Sakurai}, {Kaido}, {Horiuchi}, {Shimada}, {Inoue}, {Mitsutake},
  {Yoshida}, {Takahara}, {Takeyama}, {Suzuki}, \& {Abe}}]{2004SPIE.5487.1142I}
{Ichimoto}, K., {Tsuneta}, S., {Suematsu}, Y., {et~al.} 2004, in Presented at
  the Society of Photo-Optical Instrumentation Engineers (SPIE) Conference,
  Vol. 5487, Optical, Infrared, and Millimeter Space Telescopes. Edited by
  Mather, John C. Proceedings of the SPIE, Volume 5487, pp. 1142-1151 (2004).,
  ed. J.~C. {Mather}, 1142--1151

\bibitem[{{Keller}(1992)}]{1992Natur.359..307K}
{Keller}, C.~U. 1992, \nat, 359, 307

\bibitem[{{Kosugi} {et~al.}(2007){Kosugi}, {Matsuzaki}, {Sakao}, {Shimizu},
  {Sone}, {Tachikawa}, {Hashimoto}, {Minesugi}, {Ohnishi}, {Yamada}, {Tsuneta},
  {Hara}, {Ichimoto}, {Suematsu}, {Shimojo}, {Watanabe}, {Shimada}, {Davis},
  {Hill}, {Owens}, {Title}, {Culhane}, {Harra}, {Doschek}, \&
  {Golub}}]{2007SoPh..243....3K}
{Kosugi}, T., {Matsuzaki}, K., {Sakao}, T., {et~al.} 2007, \solphys, 243, 3

\bibitem[{{M{\"o}stl} {et~al.}(2006){M{\"o}stl}, {Hanslmeier}, {Sobotka},
  {Puschmann}, \& {Muthsam}}]{2006SoPh..237...13M}
{M{\"o}stl}, C., {Hanslmeier}, A., {Sobotka}, M., {Puschmann}, K., \&
  {Muthsam}, H.~J. 2006, \solphys, 237, 13

\bibitem[{{Muller}(1983)}]{1983SoPh...85..113M}
{Muller}, R. 1983, \solphys, 85, 113

\bibitem[{{Muller} \& {Keil}(1983)}]{1983SoPh...87..243M}
{Muller}, R. \& {Keil}, S.~L. 1983, \solphys, 87, 243

\bibitem[{{Muller} \& {Roudier}(1992)}]{1992SoPh..141...27M}
{Muller}, R. \& {Roudier}, T. 1992, \solphys, 141, 27

\bibitem[{{Muller} {et~al.}(1994){Muller}, {Roudier}, {Vigneau}, \&
  {Auffret}}]{1994A&A...283..232M}
{Muller}, R., {Roudier}, T., {Vigneau}, J., \& {Auffret}, H. 1994, \aap, 283,
  232

\bibitem[{{Osherovich} {et~al.}(1983){Osherovich}, {Chapman}, \&
  {Fla}}]{1983ApJ...268..412O}
{Osherovich}, V.~A., {Chapman}, G.~A., \& {Fla}, T. 1983, \apj, 268, 412

\bibitem[{{Parker}(1983)}]{1983ApJ...264..642P}
{Parker}, E.~N. 1983, \apj, 264, 642

\bibitem[{{Parker}(1988)}]{1988ApJ...330..474P}
{Parker}, E.~N. 1988, \apj, 330, 474

\bibitem[{{Rutten}(1999)}]{1999ASPC..184..181R}
{Rutten}, R.~J. 1999, in Astronomical Society of the Pacific Conference Series,
  Vol. 184, Third Advances in Solar Physics Euroconference: Magnetic Fields and
  Oscillations, ed. B.~{Schmieder}, A.~{Hofmann}, \& J.~{Staude}, 181--200

\bibitem[{{S{\'a}nchez Almeida} {et~al.}(2004){S{\'a}nchez Almeida},
  {M{\'a}rquez}, {Bonet}, {Dom{\'{\i}}nguez Cerde{\~n}a}, \&
  {Muller}}]{2004ApJ...609L..91S}
{S{\'a}nchez Almeida}, J., {M{\'a}rquez}, I., {Bonet}, J.~A., {Dom{\'{\i}}nguez
  Cerde{\~n}a}, I., \& {Muller}, R. 2004, \apjl, 609, L91

\bibitem[{{Sch{\"u}ssler} {et~al.}(2003){Sch{\"u}ssler}, {Shelyag},
  {Berdyugina}, {V{\"o}gler}, \& {Solanki}}]{2003ApJ...597L.173S}
{Sch{\"u}ssler}, M., {Shelyag}, S., {Berdyugina}, S., {V{\"o}gler}, A., \&
  {Solanki}, S.~K. 2003, \apjl, 597, L173

\bibitem[{{Shelyag} {et~al.}(2006){Shelyag}, {Erd{\'e}lyi}, \&
  {Thompson}}]{2006ApJ...651..576S}
{Shelyag}, S., {Erd{\'e}lyi}, R., \& {Thompson}, M.~J. 2006, \apj, 651, 576

\bibitem[{{Spruit}(1976)}]{1976SoPh...50..269S}
{Spruit}, H.~C. 1976, \solphys, 50, 269

\bibitem[{{Stanislavsky} \& {Weron}(2009)}]{2009Ap&SS.tmp..104S}
{Stanislavsky}, A. \& {Weron}, K. 2009, \apss, 104

\bibitem[{{Stanislavsky} \& {Weron}(2007)}]{2007Ap&SS.312..343S}
{Stanislavsky}, A.~A. \& {Weron}, K. 2007, \apss, 312, 343

\bibitem[{{Steiner} {et~al.}(1998){Steiner}, {Grossmann-Doerth}, {Knoelker}, \&
  {Schuessler}}]{1998ApJ...495..468S}
{Steiner}, O., {Grossmann-Doerth}, U., {Knoelker}, M., \& {Schuessler}, M.
  1998, \apj, 495, 468

\bibitem[{{Steiner} {et~al.}(2001){Steiner}, {Hauschildt}, \&
  {Bruls}}]{2001A&A...372L..13S}
{Steiner}, O., {Hauschildt}, P.~H., \& {Bruls}, J. 2001, \aap, 372, L13

\bibitem[{{Suematsu} {et~al.}(2008){Suematsu}, {Tsuneta}, {Ichimoto},
  {Shimizu}, {Otsubo}, {Katsukawa}, {Nakagiri}, {Noguchi}, {Tamura}, {Kato},
  {Hara}, {Kubo}, {Mikami}, {Saito}, {Matsushita}, {Kawaguchi}, {Nakaoji},
  {Nagae}, {Shimada}, {Takeyama}, \& {Yamamuro}}]{2008SoPh..tmp...26S}
{Suematsu}, Y., {Tsuneta}, S., {Ichimoto}, K., {et~al.} 2008, \solphys, 26

\bibitem[{{Utz} {et~al.}(2009{\natexlab{a}}){Utz}, {Hanslmeier}, {M{\"o}stl},
  {Muller}, {Veronig}, \& {Muthsam}}]{2009A&A...498..289U}
{Utz}, D., {Hanslmeier}, A., {M{\"o}stl}, C., {et~al.} 2009{\natexlab{a}},
  \aap, 498, 289

\bibitem[{{Utz} {et~al.}(2009{\natexlab{b}}){Utz}, {Hanslmeier}, {Muller},
  {Veronig}, {Muthsam}, \& {M{\"o}stl}}]{2009Hvar}
{Utz}, D., {Hanslmeier}, A., {Muller}, R., {et~al.} 2009{\natexlab{b}}, Central
  European Astrophysical Bulletin, 33, 29

\bibitem[{{Viticchi{\'e}} {et~al.}(2009){Viticchi{\'e}}, {Del Moro},
  {Berrilli}, {Bellot Rubio}, \& {Tritschler}}]{2009ApJ...700L.145V}
{Viticchi{\'e}}, B., {Del Moro}, D., {Berrilli}, F., {Bellot Rubio}, L., \&
  {Tritschler}, A. 2009, \apjl, 700, L145

\bibitem[{{V{\"o}gler} {et~al.}(2005){V{\"o}gler}, {Shelyag}, {Sch{\"u}ssler},
  {Cattaneo}, {Emonet}, \& {Linde}}]{2005A&A...429..335V}
{V{\"o}gler}, A., {Shelyag}, S., {Sch{\"u}ssler}, M., {et~al.} 2005, \aap, 429,
  335

\bibitem[{{Wiehr} {et~al.}(2004){Wiehr}, {Bovelet}, \&
  {Hirzberger}}]{2004A&A...422L..63W}
{Wiehr}, E., {Bovelet}, B., \& {Hirzberger}, J. 2004, \aap, 422, L63

\bibitem[{{Yi} \& {Engvold}(1993)}]{1993SoPh..144....1Y}
{Yi}, Z. \& {Engvold}, O. 1993, \solphys, 144, 1

\end{thebibliography}

\newpage
\begin{appendix}
\section{Correction factors for the lifetime distribution}

\stepcounter{footnote}           
\begin{table*}
	\centering
	\caption{shows the probability relations (fractional relation) between the true distribution ($N_t$; indicated by the first row) and the actually measured distribution ($N_m$; indicated by the first column)${}^{\thefootnote}$.}
		\begin{tabular}{l l l l l l l}
		\hline
			&measured&&&&&\\
			\hline
			true&$N(t)$&$N_t(0)$&$N_t(\Delta t)$&$N_t(2 \cdot \Delta t)$&$N_t(3 \cdot \Delta t)$&\dots\\
			\hline
			&$N_m(0)$&$p$&$p\cdot (1-p)$&$p\cdot (1-p)$&$p\cdot (1-p)$&\dots\\
			\hline
			&$N_m(\Delta t)$&&$p^2$&$p^2\cdot (1-p)$&$p^2\cdot (1-p)$&\dots\\
			\hline
			&$N_m(2 \cdot \Delta t)$&&&$p^3$&$p^3\cdot (1-p)$&\dots\\
			\hline
				&$N_m(3 \cdot \Delta t)$&&&&$p^4$&\dots\\
			\hline
		\end{tabular}
		\label{table1}
\end{table*}
Our true MBPs liftetime distribution is assumed to be $N_t(t)=N_0 \exp(-\lambda \cdot t)$. This distribution is measured with the outlined algorithm that has a certain detection probability $p$ to detect a feature in a single image. The elements of the principal diagonal of table \ref{table1} give us the fraction of the measured ``true'' distribution in the right measurement bins, given as:
\begin{equation}
	N_{m}=N_0\cdot \exp(-\lambda\cdot t)\cdot p^{\frac{t}{\Delta t}+1}.
	\label{equ_true}
\end{equation}
We are now deriving the higher order correction factors for the measured distribution. In Table \ref{table1} we see that there are a lot more elements. The first row (true) indicates the number of MBPs at a certain time instance $t$. The first column (measured) indicates the measured number of MBPs for a certain time instance $t$. Every element of a crossing of a row (refering to a certain measured lifetime $\mathrm{t_m}$) with a column (refering to a certain true lifetime $\mathrm{t_t}$) gives us the fraction of the number of MBPs with a measured lifetime $\mathrm{t_m}$ but having an actual lifetime of $\mathrm{t_t}$ (see also Table \ref{table1}). We see that not only the fraction of features which are measured with the right lifetimes contribute to the number of MBPs measured at a certain time but also the other elements of the row (which live longer as they are actually measured). We can conclude that the sum of the row elements contribute to the measured number at a certain time. To consider this we introduce a correction factor $K1(t)$:
\begin{equation}
	N_{m}=N_0\cdot \exp(-\lambda\cdot t)\cdot p^{\frac{t}{\Delta t}+1}+K1(t),
	\label{equ_start}
\end{equation}
where K1 is:
 \begin{eqnarray*}
    K1(0) & = & p\cdot(1-p)\cdot N_t(\Delta t)+p\cdot(1-p)\cdot N_t(2 \cdot \Delta t) \\
    &&+p\cdot(1-p)\cdot N_t(3 \cdot \Delta t) + \dots\\
    & = & p\cdot(1-p) \cdot \int_{\Delta t}^{\infty}N_0\cdot \exp(-\lambda\cdot t) dt\\
     K1(1) & = & p^2\cdot(1-p)\cdot N_t(2 \cdot \Delta t)+p^2\cdot(1-p)\cdot N_t(3 \cdot \Delta t)\\
     &&+p^2\cdot(1-p)\cdot N_t(4 \cdot \Delta t) + \dots\\
    & = & p^2\cdot(1-p) \cdot \int_{\Delta t\cdot 2}^{\infty}N_0\cdot \exp(-\lambda\cdot t) dt\\
 K1(t)   & = & p^{\frac{t}{\Delta t}+1}\cdot(1-p) \cdot \int_{t+\Delta t}^{\infty}N_t(0)\cdot \exp(-\lambda\cdot t) dt\\
 &=&N_0\cdot \exp(-\lambda\cdot (t+\Delta t))\cdot p^{\frac{t}{\Delta t}+1}\cdot \left(\frac{1-p}{\lambda}\right). \\
\end{eqnarray*}

We see that this Eq. is very similar to Eq. \ref{equ_true}. Thus we can rewrite Eq. \ref{equ_start} as
\begin{eqnarray}
	N_{m}&=&N_0\cdot \exp(-\lambda\cdot t)\cdot p^{\frac{t}{\Delta t}+1}+K1(t)\\
	N_{m}&=&N_0\cdot \exp(-\lambda\cdot t)\cdot p^{\frac{t}{\Delta t}+1}\cdot(1+\tilde{K1}),
\end{eqnarray}
where
\begin{equation}
	\tilde{K1}=\left(\frac{1-p}{\lambda}\right)\cdot \exp(-\lambda \Delta t)
\end{equation}
is always positive. This value describes the overestimation of the measurement bins of a distribution by erroneously measuring long living features in the short living bins (by loosing track of a feature). Interestingly this factor is time-independent and a constant for certain parameters (detection probability $p$, time cadence $\Delta t$ and the physical decaying parameter $\lambda$).

Now we can derive a second order correction parameter. Loosing track of the MBPs does not only lead to erroneously measured MBP lifetimes but also creates a second order lifetime distribution. This can be explained that by loosing track of the MBP, the MBP itself will not disappear. Therefore, it can happen that we will measure the same MBP again in the future. We will now try to estimate a parameter describing this circumstance.

If we move from the principle diagonal in Table \ref{table1} to the next diagonal, this diagonal would describe the probability to have an MBP in the data set that will live for one time step more than originally measured. The sum over these elements will give us all elements that live one more time step. The next diagonal describes the MBPs that live two more time steps and so on. Therefore, the diagonal elements describe the truncated part of the original distribution which forms itself a new distribution. We will derive this distribution now:
 \begin{eqnarray}
	N_{0_{new}}&=&p\cdot (1-p)\cdot N_0\cdot \exp(-\lambda \cdot \Delta t)+\\
	&&p^2\cdot(1-p)\cdot N_0\cdot \exp(-\lambda\cdot 2 \cdot \Delta t)+\dots\\
	&=&\int_{\Delta t}^{\infty}p^{\frac{t}{\Delta t}}(1-p)\cdot N_0\cdot \exp(-\lambda\cdot t) dt\\
		N_{1_{new}}&=&p\cdot (1-p)\cdot N_0\cdot \exp(-\lambda \cdot 2 \cdot \Delta t)+\\
		&&p^2\cdot(1-p)\cdot N_0\cdot \exp(-\lambda\cdot 3 \cdot \Delta t)+\dots\\
	&=&\int_{2\cdot\Delta t}^{\infty}p^{\frac{t}{\Delta t}-1}(1-p)\cdot N_0\cdot \exp(-\lambda\cdot t) dt\\
		N_{t_{new}}&=&\int_{t+\Delta t}^{\infty}p^{\frac{\tau}{\Delta t}-\frac{t}{\Delta t}}(1-p)\cdot N_0\cdot \exp(-\lambda\cdot \tau) d\tau\\
		&=&p^{\frac{-t}{\Delta t}}\cdot (1-p)\cdot N_0\int_{t+\Delta t}^{\infty}p^{\frac{\tau}{\Delta t}}\exp(-\lambda\cdot \tau) d\tau\\
	&=&\frac{p^{\frac{-t}{\Delta t}}\cdot (1-p)\cdot N_0\cdot \exp\left(\left(-\lambda\cdot+\frac{\ln p}{\Delta t}\right)\cdot (t+\Delta t)\right)}{\lambda -\frac{\ln p}{\Delta t}}\\
	&=&K2\cdot N_0\cdot \exp(-\lambda\cdot t),	
\end{eqnarray}
where $K2$ is given by:
\begin{equation}
	K2=\left(\frac{1-p}{\lambda-\frac{\ln p}{\Delta t}}\right)\cdot \exp(-\lambda \Delta t+\ln p).
\end{equation}
This factor $K2$ describes the creation of the new distribution (the truncated one), which also contributes to the measured distribution. This truncated distribution gives rise to a truncated-truncated distribution and so forth.
Finally, we arrive at the following measured distribution:
\begin{eqnarray}
	N_{m}(t)&=&N_0\cdot \exp(-\lambda \cdot t) \cdot p^{\frac{t}{\Delta t}+1}\cdot (1+\tilde{K1})\\
	&&\cdot(1+K2+K2^2+K2^3+\dots)\\
&=&\left|{\sum_{n=0;~K2<1}^{\infty}{K2^n}=\frac{1}{1-K2}}\right|\\
	&=&\frac{N_0\cdot \exp(-\lambda \cdot t) \cdot p^{\frac{t}{\Delta t}+1}\cdot (1+\tilde{K1})}{1-K2}.	
\end{eqnarray}
As a last point, we want to give the equation for the cumulated number-lifetime measurement distribution of the MBPs:
\begin{eqnarray}
	N_{m_{cum}}(t)&=&\int_{t}^{\infty}\frac{N_0\cdot \exp(-\lambda \cdot t) \cdot p^{\frac{t}{\Delta t}+1}\cdot (1+\tilde{K1})}{1-K2} dt\\
	&=&	\frac{1+\tilde{K1}}{1-K2}\cdot\frac{p}{\lambda-\frac{\ln p}{\Delta t}} \\
	 && \cdot N_0 \cdot \exp\left(\left(-\lambda_p+\frac{1}{\Delta t}~ \ln(p)\right)~t\right)\\
	&=&	K_{cum} \cdot N_0 \cdot \exp\left(\left(-\lambda+\frac{1}{\Delta t}~ \ln(p)\right)~t\right),
\end{eqnarray}
where $K_{cum}$ is the cumulated correction factor:
\begin{equation}
	K_{cum}=\frac{1+\tilde{K1}}{1-K2}\cdot\frac{p}{\lambda-\frac{\ln p}{\Delta t}}.
\end{equation}
If we compare this with the true cumulated distribution $ N_t(t)=\frac{N_0}{\lambda}\cdot \exp(-\lambda t)$, where $N_0/\lambda$ is an interesting parameter (because it describes the total amount of MBPs in the data set), we see that we have to divide our measured $N_{m_{cum}}(0)$ by the factor $K_{cum}$ and by $\lambda$:
\begin{equation}
	N_{t_{cum}}(0)=\frac{N_{m_{cum}}(0)}{K_{cum}\cdot \lambda}.
\end{equation}
\footnotetext{As an example, let us think of the real number of MBPs at time instance $2\Delta t$ ($N_t(2\Delta t)$), this number is measured (contributes) by a factor of $p^3$ reduced for the measured number $N_m(2\Delta t)$. But not only the true number of MBPs at a certain lifetime contribute to the measured value of this lifetime. If we consider the measured number of MBPs at time $\Delta t$, we see that the true number of MBPs at time $3\cdot \Delta t$ contributes with a fraction of $p^2(1-p)$ to the measured number of MBPs at a lifetime of $\Delta t$. To get a correct distribution all of these relations have to be considered (see text).}

\end{appendix}

\end{document}